\documentclass[a4paper]{panl}
\usepackage{cite}
\usepackage{wrapfig}
\usepackage{graphicx}
\graphicspath{ {./images/} }
\DeclareGraphicsExtensions{.pdf}
\usepackage{amssymb}
\usepackage{amsfonts}
\usepackage{color}
\usepackage{amsmath}
\usepackage{longtable}
\usepackage{rotating}
\usepackage{lscape}
\usepackage{epsfig}
\usepackage{multirow}
\usepackage[hidelinks]{hyperref}
\usepackage[justification=centering]{caption}
\usepackage{lineno}
\usepackage{footnote}
\usepackage{subcaption}
\makesavenoteenv{tabular}
\makesavenoteenv{table}
\usepackage{xcolor}
\hypersetup{
	colorlinks,
	linkcolor={red!80!black},
	citecolor={green!80!black},
	urlcolor={blue!80!black}
}

\originalTeX
\begin{document}
% Journal sections (see http://pkp.jinr.ru/index.php/PEPAN_LETTERS/about/editorialPolicies#focusAndScope)
\issuearea{Physics of Elementary Particles and Atomic Nuclei. Theory}
% or in Russian
%\issuearea{ФИЗИКА ЭЛЕМЕНТАРНЫХ ЧАСТИЦ И АТОМНОГО ЯДРА. ТЕОРИЯ}

\title{On the study of antiprotons yield in hadronic collisions at NICA SPD   \\ \ Об изучении выхода антипротонов в адронных столкновениях на NICA SPD}
\maketitle
\authors{V.\,Alexakhin$^{a,}$\footnote{E-mail: vadim.alexakhin@cern.ch}, A.\,Guskov$^{a,}$\footnote{E-mail: alexey.guskov@cern.ch}, Z.\,Hayman$^{b,}$\footnote{E-mail: zmhayman@cu.edu.eg},
R.\,El-Kholy$^{b,}$\footnote{E-mail: relkholy@sci.cu.edu.eg}, A.\,Tkachenko$^{a,}$\footnote{E-mail: avt@jinr.ru}}
\from{$^{~a}$\,Joint Institute for Nuclear Research (JINR), Dubna, Russia, 141980}
\vspace{-3mm}
\from{$^{~b}$\,Astronomy Department, Faculty of Science, Cairo University, Giza, Egypt, 12613}

\begin{abstract}
The study of antiproton yield in $p$-$p$ and $p$-$d$ collisions is important for the astrophysical search for dark matter consisting of Weakly Interacting Massive Particles. Refinement of the production cross section, angular and momentum spectra of produced antiprotons in a wide energy range could help to treat the results obtained by the AMS-02 and PAMELA orbit spectrometers. In this paper we present a detailed Monte Carlo study of possible measurements at the planned SPD experiment at the NICA collider that is under construction at JINR (Dubna).

\vspace{0.42cm}

Изучение выхода антипротонов в $p$-$p$ и $p$-$d$ столкновениях является важным для астрофизических поисков тёмной материи, состоящей из слабо взаимодействующих массивных частиц (WIMP). Уточнение сечений рождения антипротонов, их угловых и импульсных спектров в широком диапазоне энергий могло бы помочь в интерпретации результатов, полученных на орбитальных спектрометрах AMS-02 и PAMELA. В данной работе мы представляем детальное Монте Карло моделирование возможных измерений в планируемом эксперименте SPD на коллайдере NICA, который строится в ОИЯИ (Дубна).
\end{abstract}
\vspace*{6pt}


\label{sec:intro}
\section*{\textsc{Introduction}}~
Dark Matter (DM) is thought to compose more than 24\% of the matter-energy content of the Universe\cite{Hinshaw2013}. While observations have not yet revealed the origin or nature of DM, there are many theories trying to infer both from available information. One of the most common theories is that the galactic dark halos are built up of thermal relics of the Big Bang called Weakly Interacting Massive Particles (WIMPs)\cite{Tanabashi2018}. There is a possibility that, even though DM particles do not interact with Standard \mbox{Model (SM)} particles, they do produce SM particles via pair-annihilation or decay. Powerful experiments, such as AMS-02 and PAMELA, have been trying to detect anomalies in rare components of Cosmic Rays (CRs), especially antiparticles like antiprotons. However, conclusions are hindered by uncertainties.

To be able to detect anomalous components in CRs, it is crucial to measure the production of these components from conventional astrophysical sources with the maximal possible precision. For the secondary antiproton background, the sources are proton-nucleus, nucleus-proton, and nucleus-nucleus collisions of CRs on the Interstellar Medium (ISM). However, almost all secondary antiprotons are a result of collisions of light nuclei. Despite the AMS-02 measuring the antiproton flux with unprecedented accuracy of a few \mbox{percents\cite{Aguilar2016}}, no conclusion can be made yet about exotic components due to several sources of uncertainty. The largest source of uncertainties, which can vary from 20 to 50\% depending on the energy, is uncertainties on antiproton-production cross sections\cite{Giesen2015}.

In addition to the lack of data of antiproton production in collisions involving helium, data of antiproton production in proton-proton collisions are scarce. Moreover, most of the existing datasets are not corrected for antiproton production via the decay of intermediate hyperons; namely, $\bar{\Lambda}$ and $\bar{\Sigma}^{-}$; a process that gains significance with increasing energy\cite{Winkler2017}. In order to keep up with the AMS-02 accuracy and energy range, new accurate measurements of antiproton production are essential. This would enable comparison of data with theoretical models; and, eventually, coming to a conclusion about any exotic signals. These measurements need to cover the proton-beam kinetic-energy range from 1 GeV to 6 TeV, and the pseudorapidity range from 2 to 8\cite{Donato2017}.

It is currently planned to measure the antiproton-production cross-sections of $p$-$p$ and $p$-$He$ collisions in fixed-target mode by the COMPASS Collaboration\cite{Adams2018}. However, the $4\pi$-universality of the Spin Physics Detector (SPD) detector\cite{Savin2015} at the Nuclotron-based Ion Collider fAcility (NICA)\cite{Kekelidze2017} will enable measurements of antiprotons with high transverse momentum ($p_{T}$) with lower beam energy. There is also the possibility of performing proton-deuteron collision measurements, which contribute to the secondary antiproton flux in CRs as well.

NICA is currently under construction at JINR. It is planned to operate the NICA collider with polarized proton and deuteron beams for spin physics study at the SPD detector. In a previous work\cite{Guskov2019}, we performed a preliminary study of antiproton production at the SPD. Monte Carlo simulations generated with Pythia8 have shown that the momenta of the antiprotons produced in $p$-$p$ collisions, at a center-of-mass (CM) energy $\sqrt{s} = 26$ Gev, would be low enough to identify with the time-of-flight method. They have also shown that with a time resolution of about 100 ps and a flight base of 2 meters, $K^{-} / \bar{p}$ separation can be achieved for antiprotons with momenta up to 4 GeV, which would include most of the antiproton yield.

In this paper, we study with more detail the antiproton production in hadronic collisions with Monte Carlo simulations specific to the SPD setup using the SPDROOT toolkit.

\break

\section{\textsc{The Spin Physics Detector}}
The Spin Physics Detector is planned as a multipurpose  universal  4$\pi$ detector. It will be able to operate in the polarized $p$-$p$ collision mode with CM energy up to \mbox{27 GeV} and luminosity up to about $10^{32}$ cm$^{-2}$ s$^{-1}$. Another mode would be the polarized $d$-$d$ collisions at energy up to 13.5 GeV and a luminosity one order lower. Asymmetric $p$-$d$ collisions are also discussed\cite{Savin2015}. The detector consists of the  barrel part and two end-caps. The tracking system of the SPD  includes the silicon-based vertex detector surrounded by the main tracker, using gas-filled drift straw-tubes as the basic detection element with a spatial resolution of about 100 $\mu$m. The magnetic field (a few configurations are currently under consideration) provides charged-particle momentum measurement with a typical resolution of about 2\% at a transverse momentum of 1 GeV/c.
The time-of-flight (TOF) system, with a time resolution below 100 ps, provides identification of secondary hadrons in a wide kinematic range. 
The shashlyk-type electromagnetic calorimeter is responsible for the photon reconstruction and identification of electrons and positrons. 
The muon system (RS) performs the advanced muon/hadron separation via comprehensive pattern recognition and matching of the track segments to the tracks in the inner part of the detector\cite{Savin2015}.

This Monte Carlo study simulation was performed in order to estimate the capability of the Spin Physics Detector for the precision measurement of antiproton yield in hadronic collisions at NICA. The study  was done using the SPDROOT toolkit, a ROOT- and GEANT4-based software developed for SPD using the FAIRROOT framework\cite{AlTurany2012}. A simplified description of SPD  with a quasi-solenoidal magnetic field along the beam axis ($B = 0.4~$T) was used. The time resolution of the TOF system used for antiproton identification was supposed to be 70 ps. PYTHIA8\cite{Sjoestrand2015} was used as a generator of primary $p$-$p$ interactions. Since the geometry of the detector is not yet fixed all the results presented in the paper should be treated as indicative.

\section{\textsc{Direct Production of Antiprotons in $p$-$p$ Collisions}}
\begin{figure}[!p]
	\centering
	\includegraphics[width=\linewidth,height=0.45\textheight]{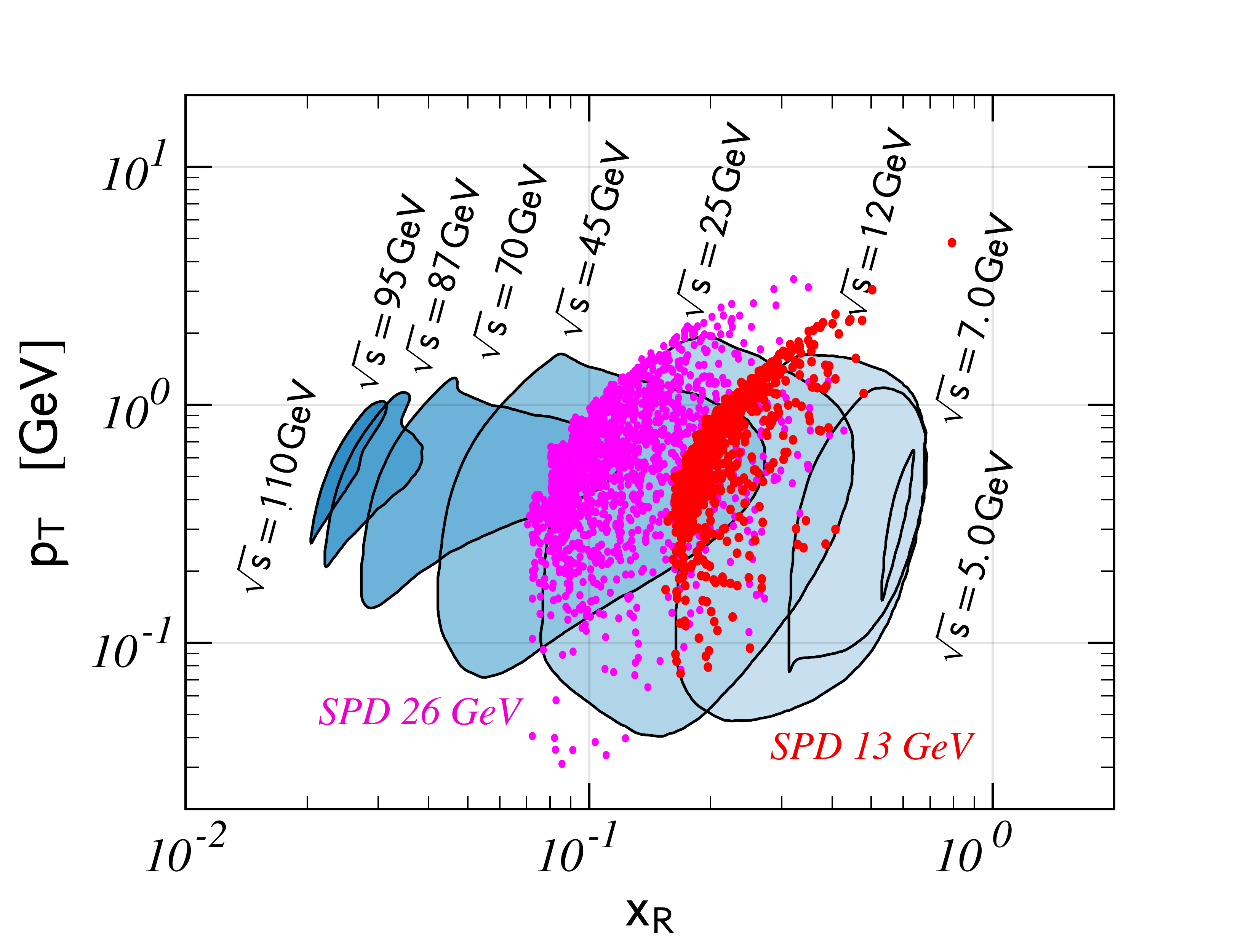}
	\caption{The kinematic range accessible by $p$-$p$ collisions at the SPD detector, where the red points represent a CM energy of 13 GeV  and the magenta points represent a CM energy of 26 GeV, superimposed on the required measurement range to match the precision level of the AMS-02 measurements\cite{Donato2017}.}
	\label{fig:NICA_range}
\end{figure}

\begin{figure}[!p]
	\centering
	\begin{subfigure}[t]{0.495\textwidth}
		\centering
		\includegraphics[width=\linewidth, height=0.3\textheight]{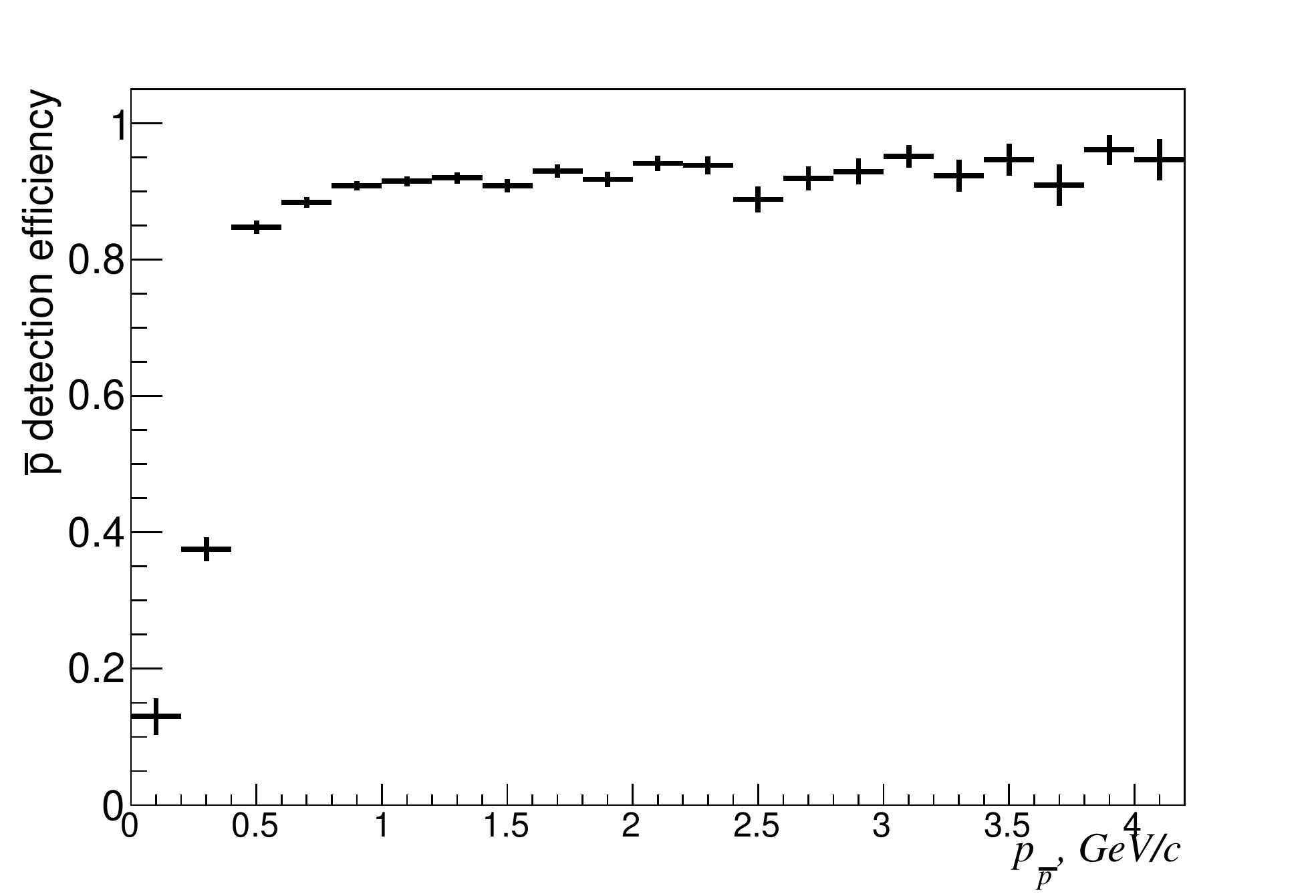}
		\caption{~}
		\label{fig:efficiency}
	\end{subfigure}
	\hfill
	\begin{subfigure}[t]{0.495\textwidth}
		\centering
		\includegraphics[width=\linewidth, height=0.3\textheight]{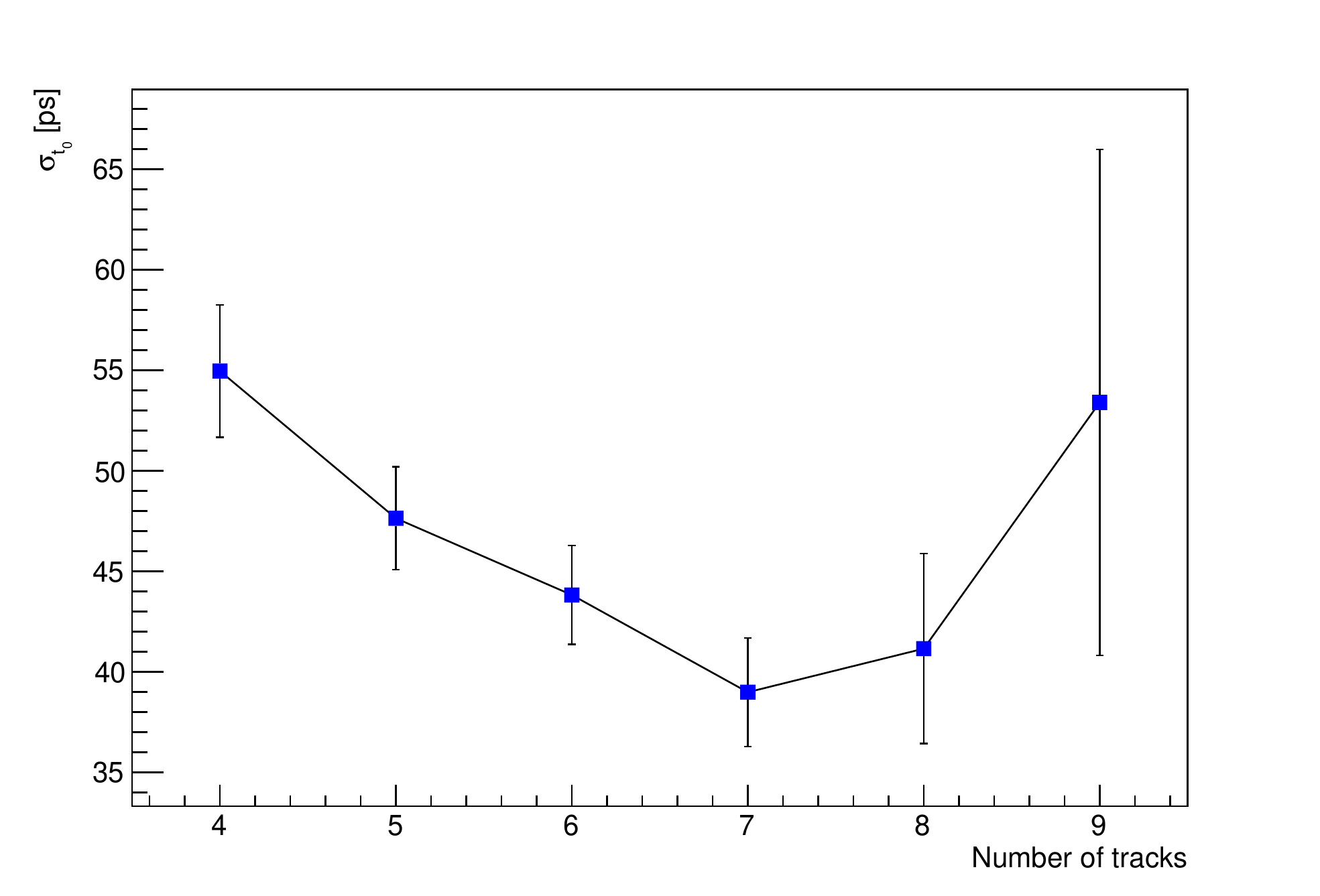}
		\caption{~}
		\label{fig:resolution}
	\end{subfigure}
	\caption{(a) Detection efficiency as a function of $\bar{p}$ momentum for $p$-$p$ collisions at  $\sqrt{s}=26$ GeV. (b) $t_0$ reconstruction accuracy as a function of the number of charged tracks in the event. }
	%	\label{fig:efficiency-resolution}
\end{figure}

The antiproton-production cross section in $p$-$p$ collisions multiplied by the average antiproton-multiplicity in the discussed kinematic range is on the level of a few \mbox{millibarns \cite{Guskov2019}}. That corresponds to an antiproton production rate on a level $>10^5$ s$^{-1}$.  So, the accuracy of the proposed measurements is not limited by the statistical uncertainty. The phase space in terms of the variable $x_R$ and the transverse momentum $p_T$ of produced antiprotons covered by SPD at $\sqrt{s}=13$ GeV and $\sqrt{s}=26$ GeV is shown in Fig.\ref{fig:NICA_range}. Here $x_R$ is defined as the ratio of antiproton energy to its maximal possible energy for a particular $\sqrt{s}$. The geometrical acceptance of the tracking system that affects the track reconstruction efficiency and energy losses of the produced antiprotons, and the magnetic field preventing low-energy particles from reaching the TOF system are taken into account. An antiproton produced at the interaction point was considered detected if it reached the TOF system in the barrel or an end-cap. The radius of the barrel part of the TOF system and the distance from the center of the detector to the end-cap parts are assumed to be 2.0 m each.
The detection efficiency as a function of $\bar{p}$ momentum for $\sqrt{s}=26$ GeV is presented in Fig.\ref{fig:efficiency}. 

\begin{figure}[!h]
	\includegraphics[width=1.1\textwidth]{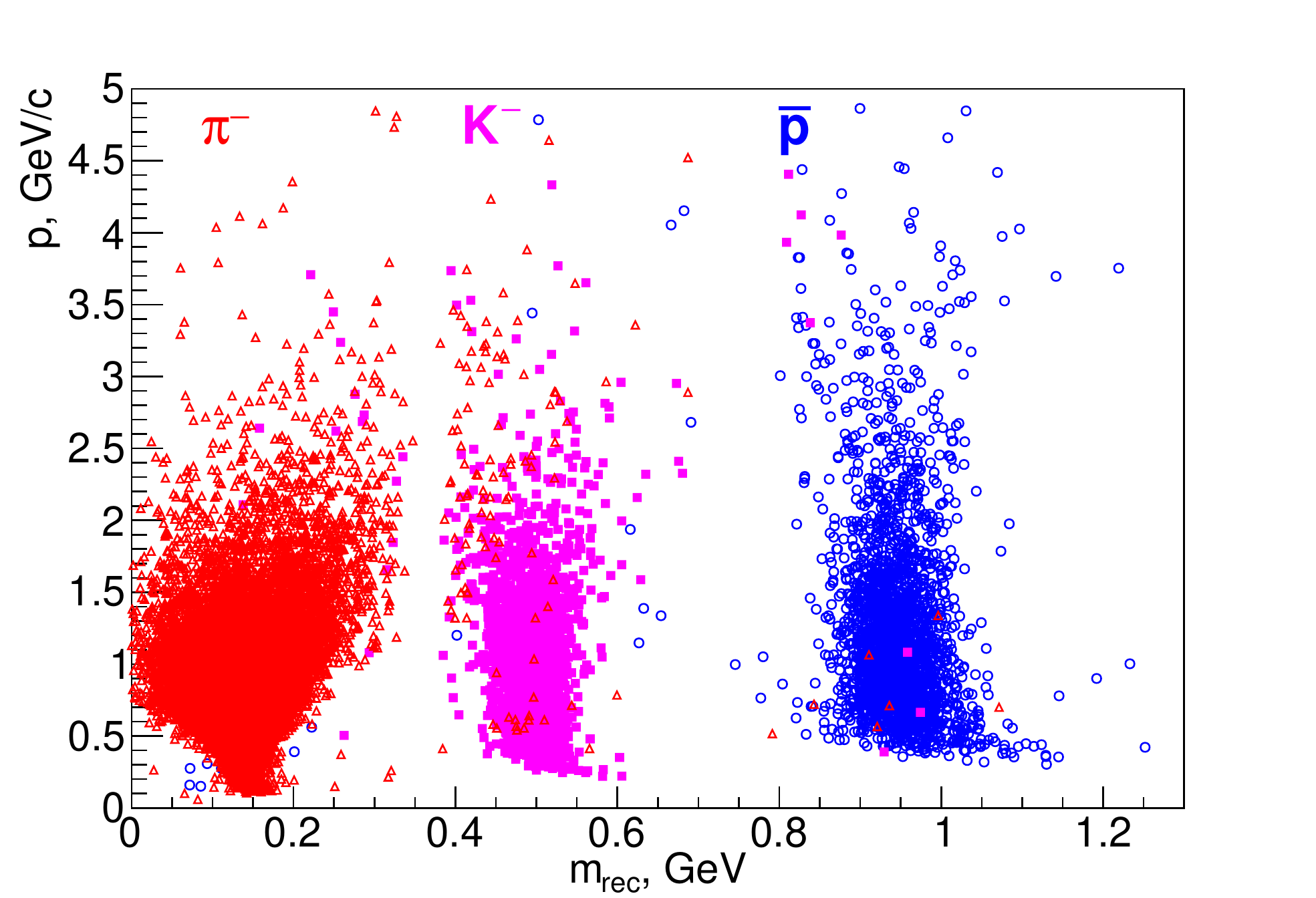}
	\caption{Mass reconstruction for $\pi^{-}$, $K^{-}$, and $\bar{p}$ using the TOF system.\\Simulation is performed for $\sqrt{s}=26$ GeV.}
	\label{fig:mass-momentum}
\end{figure}

Antiproton identification is the most critical issue for the proposed measurements. The situation is worsening with the presence of only one detector plane for the time-of-flight measurement and with the relatively-low multiplicity of secondary charged particles. The algorithm described in \mbox{Ref. \cite{Adam2017}} was applied for events with track multiplicities \mbox{above 3}. For each track in the event, three possible hypotheses were assumed: $\pi$, $K$ and $p$. The $\chi^2$ minimization of the time mismatch at the interaction point was performed over all possible combinations of individual-track hypotheses. As a result of such minimization, for each event the interaction time $t_0$ and the most probable particle IDs for each track could be found. Fig.\ref{fig:resolution} shows the accuracy of $t_0$ reconstruction as a function of the number of charged tracks in the event. Particle-mass reconstruction using TOF system is illustrated in Fig.\ref{fig:mass-momentum}. Safe $K^{-}/\bar{p}$ separation with a purity of antiproton sample on the level of 1\% could be provided for particles with momenta below 3.5 GeV/$c$.
\vspace{-0.45cm}
\pagebreak
\section{\textsc{Antiprotons from Anti-Hyperons Decay}}
\begin{figure}[!h]
	\centering
	\includegraphics[width=\linewidth,height=0.39\textheight]{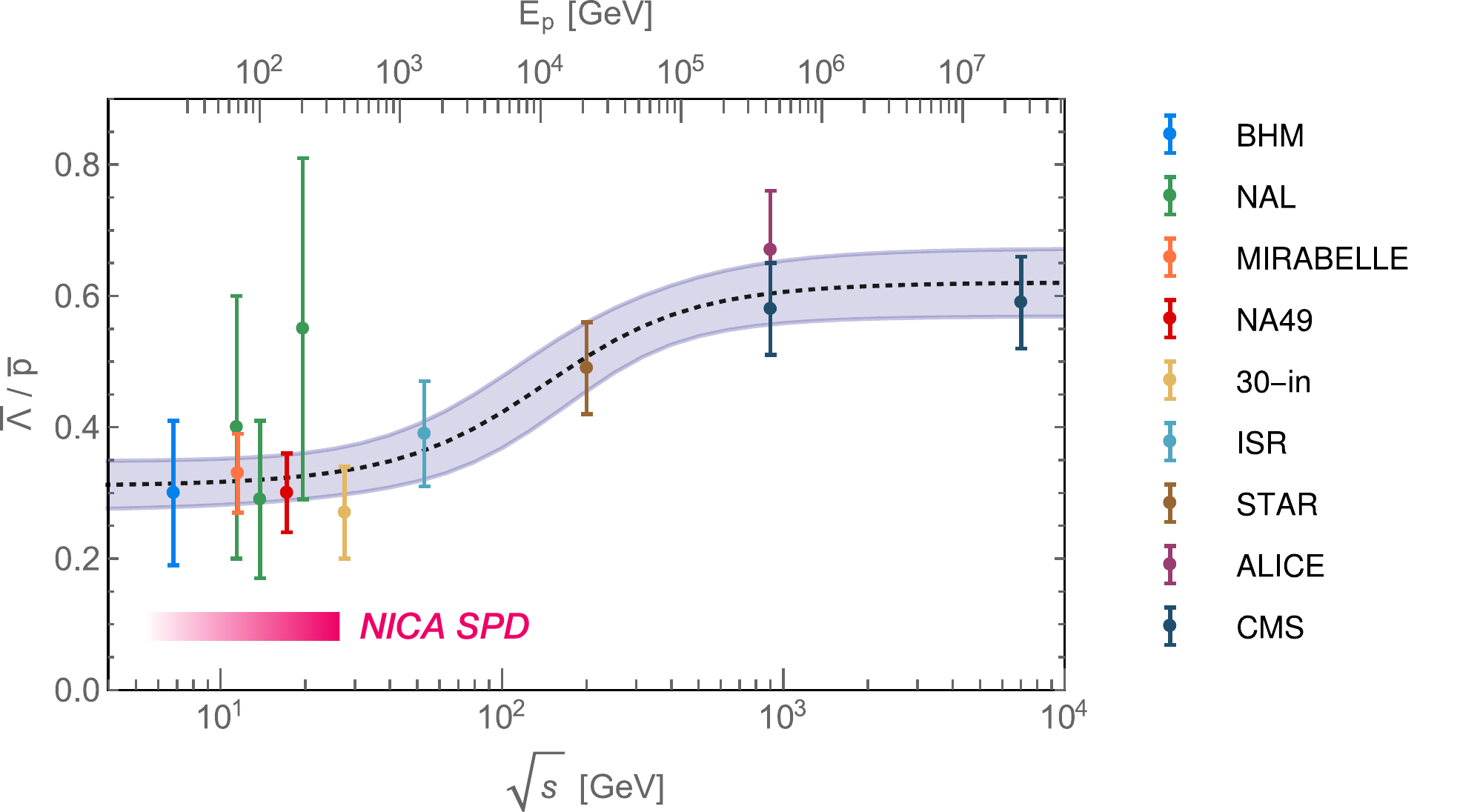}
	\caption{$\bar{\Lambda}/\bar{p}$  ratio in proton-proton collisions as measured by several experiments \cite{Winkler2017}. The range of $\sqrt{s}$ accessible at the NICA SPD is shown in red at the lower left part.}
	\label{fig:LambdaR}
\end{figure}

It has been mentioned in \cite{Guskov2019} that a significant part of antiprotons is produced from the decays of secondary $\bar{\Lambda}$ and $\bar{\Sigma}^{-}$ anti-hyperons. The ratio $\Delta_{\Lambda}$ of hyperon-induced to directly produced antiprotons can be expressed as
\begin{equation}
  \Delta_{\Lambda} = \frac{\bar{\Lambda}}{\bar{p}} \times \text{BF}_{\bar{\Lambda}\to \bar{p} \pi^+}
  +  \frac{\bar{\Sigma}^-}{\bar{p}} \times \text{BF}_{\bar{\Sigma}^-\to \bar{p} \pi^0},
\end{equation}
where $\bar{\Lambda}/\bar{p}$ and ${\bar{\Sigma}}^-/\bar{p}$ are the hyperon to (promptly produced) antiproton ratios, $\text{BF}_{\bar{\Lambda}\to \bar{p} \pi^+}$ $= 0.639\pm0.005$ and $\text{BF}_{\bar{\Sigma}^-\to \bar{p} \pi^0}=0.5157\pm0.0003$ are the branching fractions of the corresponding decays. Existing experimental data for the ratio $\bar{\Lambda}/\bar{p}$ are presented in Fig.\ref{fig:LambdaR}. The relative uncertainty of this quantity at SPD energies is about  12\%. As for the $\bar{\Sigma}^{-}/\bar{p}$  ratio, there is not any trustable experimental data on it; and it was estimated roughly based on the symmetry arguments that $\bar{\Sigma}/\bar{\Lambda}=0.33$ \cite{Kappl2014}. Thus, $\Delta_{\Lambda} = (0.81\pm0.04)\times \bar{\Lambda}/\bar{p}$.

\begin{figure}[!h]
	\centering
	\begin{subfigure}[t]{0.495\textwidth}
		\centering
		\includegraphics[width=\linewidth, height=0.3\textheight]{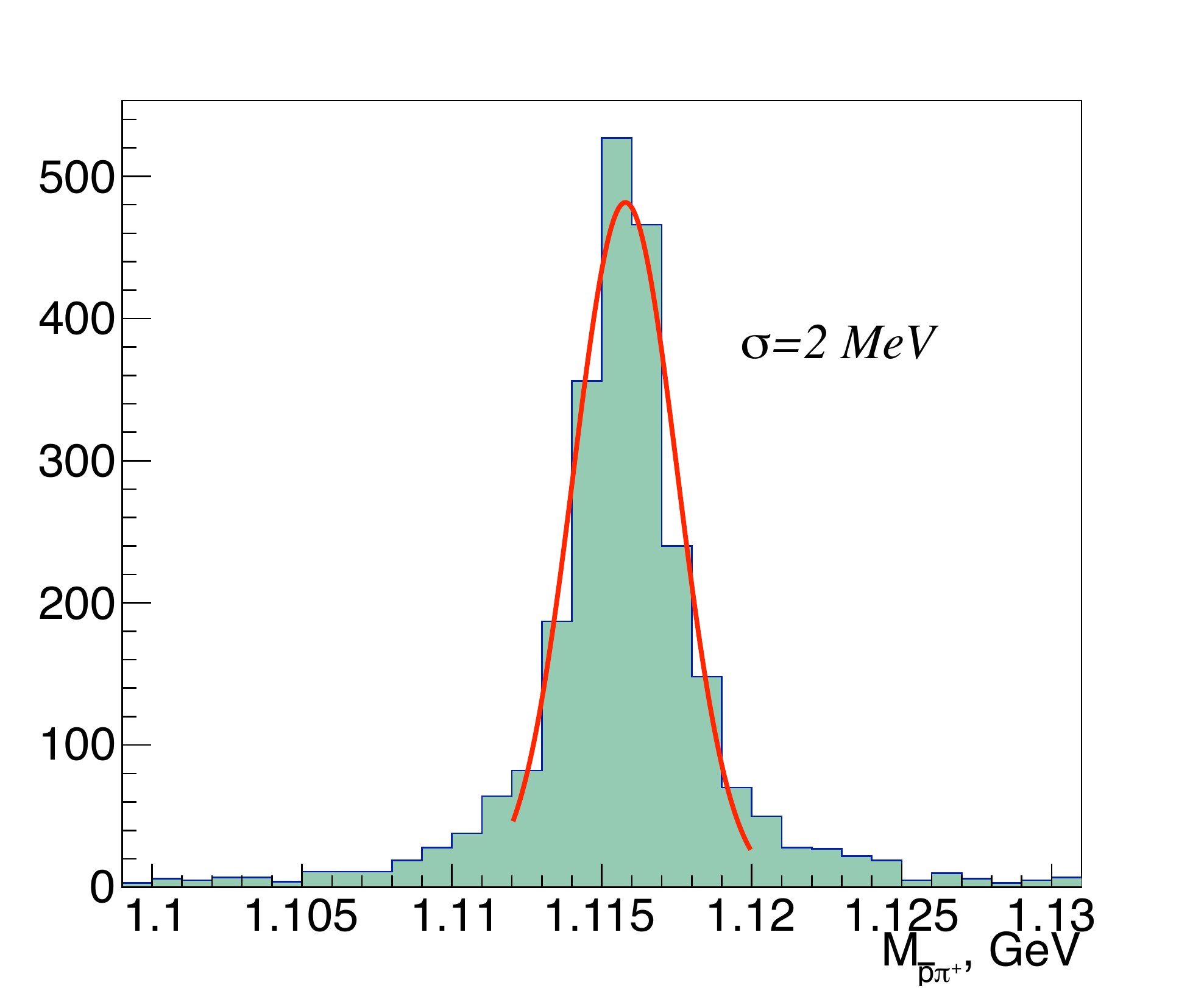}
		\caption{~}
		\label{fig:Lambda_a}
	\end{subfigure}
	\hfill
	\begin{subfigure}[t]{0.495\textwidth}
		\centering
		\includegraphics[width=\linewidth, height=0.3\textheight]{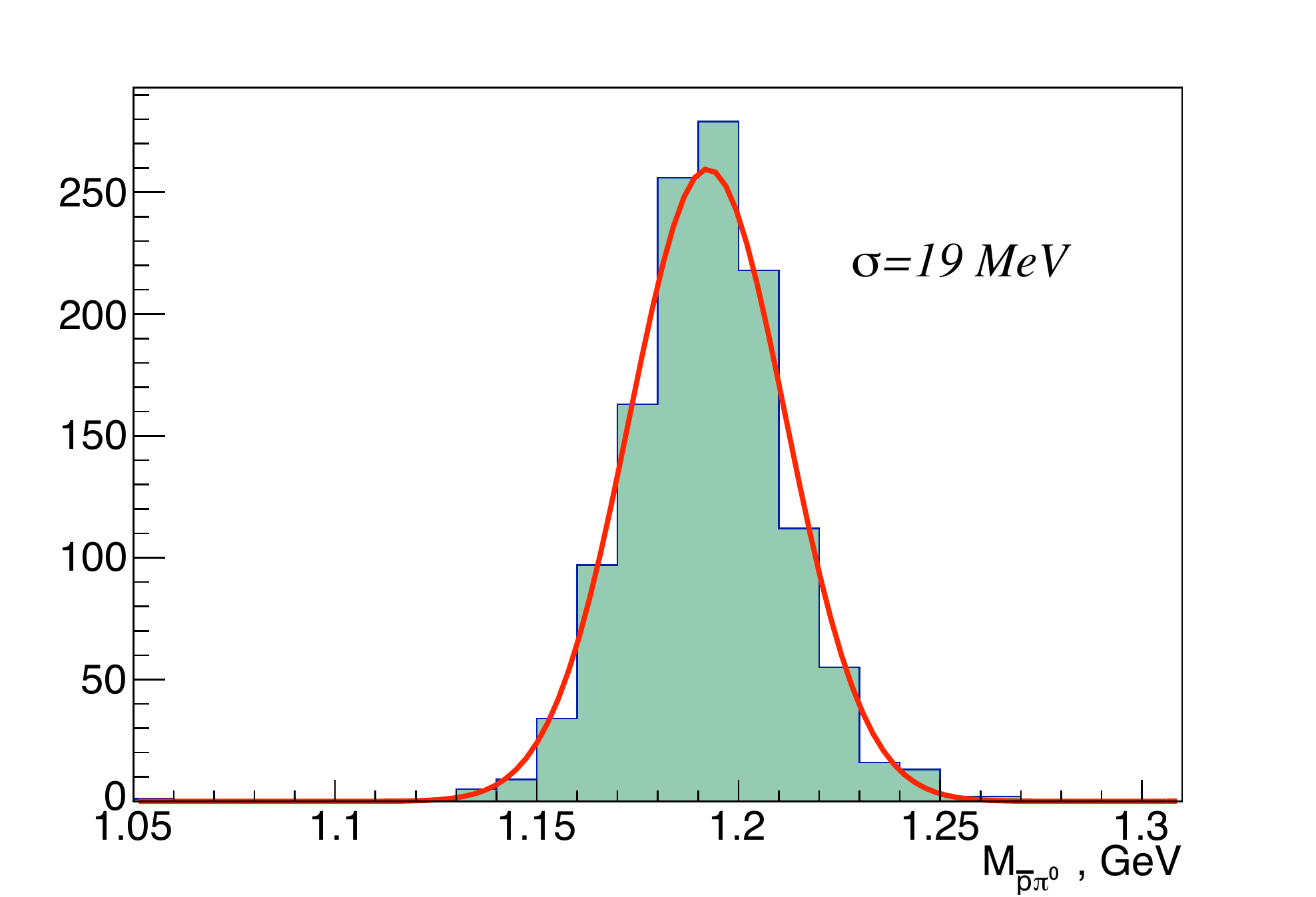}
		\caption{~}
		\label{fig:Lambda_b}
	\end{subfigure}
	\caption{(a) Reconstructed  $\bar{p}\pi^{+}$ invariant mass spectrum with a $\bar{\Lambda}$ signal.\\(b) Reconstructed $\bar{p}\pi^{0}$ invariant mass spectrum with a $\bar{\Sigma}^-$ signal.  }
\end{figure}

Since hyperons have a macroscopic decay length, their contribution can be accessed at the SPD via reconstruction of the two-body decays $\bar{\Lambda}\to \bar{p}\pi^+$ and $\bar{\Sigma}^-\to \bar{p}\pi^0$ in secondary vertices. The fraction of $\bar{\Lambda}$ decaying within the inner part of the SPD setup, averaged over the spectrum, is 93\% at $\sqrt{s}=26~GeV$ while almost all  $\bar{\Sigma}^-$ decay there. The reconstructed invariant mass of  $\bar{p}\pi^{+}$ from $\bar{\Lambda}$ decay in secondary V-vertices is shown in Fig.\ref{fig:Lambda_a}. The Gaussian width of the peak is about 2 GeV/$c^2$. It is fully defined by the uncertainty of the momentum measurement in the SPD tracking system. 

For determination of the $\bar{\Sigma}^-$ decay, a secondary vertex should be reconstructed using a track pointing to the primary vertex and associated with $\bar{\Sigma}^-$ and a track associated with an antiproton. A pair of photons detected in the electromagnetic calorimeter should be assumed to be produced from a $\pi^0$ decay in the found secondary vertex. The width of the reconstructed $\bar{\Sigma}^-$ peak is mainly defined by the energy resolution of the calorimeter and equals to 19 MeV as it can be seen from Fig.\ref{fig:Lambda_b}. Both signals, $\bar{\Lambda}$ and $\bar{\Sigma}^-$, are well-distinguished over the background. That inspires a hope to determine the ratios $\bar{\Lambda}/\bar{p}$ and ${\bar{\Sigma}}^-/\bar{p}$ and, finally, $\Delta_{\Lambda}$ with a minimal statistical uncertainty.

\section{\textsc{Summary}}
Based on the proposal to use the Spin Physics Detector at the NICA collider as an instrument for the precision measurements of the antiproton yield in $p$-$p$ and $p$-$d$ collisions required by astrophysical searches for dark matter \cite{Guskov2019}, we performed this Monte Carlo study using a preliminary SPD setup geometry. We refined the kinematic coverage of the detector, studied its possibility to identify directly produced antiprotons with the TOF system, and investigated the conditions of reconstructing the $\bar{\Lambda}$ and $\bar{\Sigma}^{-}$ hyperon decays. It was shown that the SPD TOF system on the base of 2.0 m can provide purity of $K^{-}/\bar{p}$ separation better than 99\% up to $p=3.5$ GeV/$c$. Accuracy of the interaction time $t_0$ reconstruction could reach 40-50 ps. The SPD tracking system and electromagnetic calorimeter offer the possibility of effectively reconstructing the decays $\bar{\Lambda} \to \bar{p}\pi^{+}$ and $\bar{\Sigma}^-\to \bar{p} \pi^0$, which are the main sources of secondary antiprotons. The setup resolution of the $\bar{\Lambda}$ and $\bar{\Sigma}^-$ peaks is found to be 2 MeV and 19 MeV, respectively.

We reaffirm our previous conclusions that the supplementary measurements at the  Spin Physics Detector could make a sizable contribution to the search of physics beyond the Standard Model.

\bibliographystyle{pepan}
\bibliography{ref}

\end{document}